\documentclass[aps,prc,twocolumn,superscriptaddress,showpacs]{revtex4-1}

\usepackage[dvips]{graphicx}
\usepackage{amsmath,amssymb}
\usepackage[T1]{fontenc}

\begin{document}

\title{Neutrino absorption by hot nuclei in supernova environments}

\author{Alan~A.~Dzhioev}
\email{dzhioev@theor.jinr.ru}
\affiliation{Bogoliubov Laboratory of Theoretical Physics, JINR, 141980, Dubna, Russia}
\author{A.~I.~Vdovin}
\email{vdovin@theor.jinr.ru}
\affiliation{Bogoliubov Laboratory of Theoretical Physics, JINR, 141980, Dubna, Russia}
 \author{J.~Wambach}
 \email{jochen.wambach@physik.tu-darmstadt.de}
\affiliation{Institut f{\"u}r Kernphysik, Technische Universit{\"a}t Darmstadt, 64289 Darmstadt, Germany}
\affiliation{GSI Helmholtzzentrum f\"ur Schwerionenforschung, Planckstr. 1, 64291 Darmstadt, Germany}

\date{\today}

\begin{abstract}

Using the thermal quasiparticle random phase approximation, we study the process of neutrino and antineutrino capture on hot nuclei in supernova environments.
For the sample nuclei $^{56}$Fe and $^{82}$Ge we perform a detailed analysis of thermal effects on the strength distribution of allowed Gamow-Teller transitions
which dominate low-energy charged-current neutrino reactions. The finite temperature cross sections are calculated taking into account the contributions of
both allowed and forbidden transitions. The enhancement of the low-energy cross sections is explained by considering thermal effects on the GT$_\pm$ strength.
For $^{56}$Fe we compare the calculated finite-temperature cross sections with those obtained from large-scale shell-model calculations.

\end{abstract}


% see http://www.aip.org/publishing/pacs/pacs-alphabetical-index

% 26.50.+x Nuclear physics aspects of novae, supernovae, and other explosive environments
% 21.60.Jz Random-phase approximation (nuclear structure)
% 24.10.Pa Statistical models of nuclear reactions
% 25.30.Pt Neutrino-nucleus scattering

\maketitle

\section{Introduction}

It is well known that core-collapse supernova simulations require a detailed description of neutrino transport including all potentially important neutrino reactions~\cite{Janka_PhysRep442}. Neutrinos are the mediators of the energy transfer from the core to the outer stellar layers and their luminosities and spectra are a crucial ingredient for the supernova explosion mechanism. Despite significant progress in our understanding  of the core-collapse mechanism, many supernova simulations fail to produce explosion (see e.g.  \cite{Liebendoerfer,Mezzacappa} and references therein). One of the possible reasons for this failure could be the incomplete and inaccurate treatment of the neutrino-nucleus processes occurring in supernova environments.

It was pointed by Haxton~\cite{Haxton_PRL60} that neutral- and charged-current neutrino reactions on nuclei involving the excitation of the giant resonances might help reenergize the explosion and they should, therefore, be taken into account in supernova simulations. Such simulations were performed by Bruenn and Haxton~\cite{Bruenn_APJ376}
and it was shown that inelastic neutrino-nucleus scattering plays an important role in equilibrating neutrinos with matter. However, the simulations did not confirm Haxton's suggestion that neutrino-nucleus reactions might help to revive the stalled shock wave.

In~\cite{Bruenn_APJ376}, the nuclear composition of the core was approximated by a single representative nucleus, $^{56}$Fe, and the relevant cross sections and rates were evaluated considering only allowed (Gamow-Teller) and first-forbidden upward transitions from the nuclear ground state.  However, in the hot supernova environment with temperatures $T\gtrsim1$~MeV nuclei exist as a thermal ensemble, i.e.,
nuclear excited states are  thermally populated according to the Boltzmann distribution. As was first realized in Ref.~\cite{Fuller_APJ376}, downward transitions from excited states remove the reaction threshold and can significantly enhance the reaction cross section at low neutrino energies.

From  a microscopic point of view, there are two ways how to treat neutrino reactions with hot nuclei. The first one involves a state-by-state summation over Boltzmann-weighted, individually determined contributions from nuclear ground and excited states. To apply this method, one needs to know the strength distribution of electro-weak transition operators  for thermally populated nuclear states. The second method is based on a statistical formulation  of the nuclear many-body problem. In this approach rather than computing individual strength distributions one  determines  an "average" temperature-dependent strength function.

For iron group nuclei ($A=45-65$) the first approach was developed in~\cite{Sampaio_PLett511,Sampaio_PLetB529,Juodagalvis_NPA747} on the basis of large-scale shell-model (LSSM) diagonalization calculations. Modern high-performance compute capabilities combined with state-of-the-art diagonalization approaches are able to provide detailed strength distributions for both charge-neutral and charge-changing Gamow-Teller transitions that dominate neutrino reactions with $pf$-shell iron group nuclei at low neutrino energies ($E_\nu\lesssim 20~\mathrm{MeV}$). However, at temperatures $T\gtrsim1~\mathrm{MeV}$ an explicit state-by-state summation over all thermally populated states presently remains computationally prohibitive. To overcome this problem the "Brink hypothesis" is applied, i.e., it is assumed that GT distributions on nuclear excited states are the same as for the ground state.  To account for thermal effects, the giant resonances built on the ground and low-lying daughter states in the inverse reaction are considered. Such excited states are called back-resonances and their importance arises from the large nuclear matrix elements and increased phase space.
Thus, within LSSM calculations the temperature dependence of the cross sections is comprised solely in the back-resonance terms.

Within the shell model approach it is assumed that the GT distributions on thermally excited states are the same as for the ground state. This is not likely the case as the vanishing of pairing correlations and smearing of the Fermi surface with increasing temperature should affect the distribution centroid and move it slightly down in energy. This conjecture is confirmed by shell model Monte Carlo studies performed at finite temperature~\cite{Radha_PRC56}. In addition, with present compute capabilities the shell-model diagonalization cannot be applied for nuclei beyond the $pf$-shell because of the huge configurational spacse involved. These shortcomings can be avoided in  a thermal quasiparticle random-phase approximation (TQRPA). In~\cite{Dzhioev_IJMPE18}, the TQRPA was proposed as a method to study the response of hot nuclei to weak external perturbations in the framework of a statistical approach. Being based on the thermo-field dynamics (TFD) formalism~\cite{Takahashi_IJMPB10,Umezawa1982,Ojima_AnPhys137}, the TQRPA enables the computation of temperature-dependent strength function  avoiding the assumption of Brink's hypothesis.
In~\cite{Dzhioev_PRC89,Dzhioev_PAN74}, the TQRPA was applied to study thermal effects on neutral-current inelastic neutrino scattering on $^{54,56}$Fe and neutron-rich germanium isotope $^{82}$Ge. It was shown that the TQRPA reveals the same thermal effects on the cross sections as the LSSM approach. Namely, the reaction threshold for inelastic neutrino-nucleus scattering is removed at finite temperature and the cross section for low-energy neutrinos is significantly enhanced. It was found, however, that within the TQRPA the enhancement is caused by both de-excitation of nuclear excited states and thermally unblocked low-energy GT transitions. The latter do not appear within the LSSM due to application  of Brink's hypothesis. Moreover, it was shown in~\cite{Dzhioev_PRC89} that, unlike in the LSSM approach, the principle of detailed balance is not violated within the TQRPA and it results in a larger strength for downward GT transitions from excited states.  As a consequence, at low neutrino energies the finite temperature cross sections calculated within the TQRPA turn out to be several times larger than those obtained within the shell-model calculations.

In the present paper, we apply the TQRPA method to study thermal effects on charged-current neutrino-nucleus reactions $(\nu_e,e^-)$ and $(\overline{\nu}_e,e^+)$ occurring in the supernova environment. Here we would like to mention that the TQRPA  was already applied for charge-changing reactions with hot nuclei
when studying thermal effects on electron capture in supernovae~\cite{Dzhioev_PAN72,Dzhioev_PRC81}.
The paper is organized as follows: in Sec.~\ref{formalism} we review the basics of the TFD formalism and outline how to treat charge-changing transitions in a hot nucleus within the TQRPA. For more details of the approach, the reader is referred to Refs.~\cite{Dzhioev_PAN72,Dzhioev_PRC81}. In addition, in Sec.~\ref{formalism} the expressions necessary to calculate cross-sections for $\nu_e$- and $\overline{\nu}_e$-absorption on hot nuclei are given. The results of the numerical calculations are presented and discussed in Sec.~\ref{results} for
the sample nuclei $^{56}$Fe and $^{82}$Ge. For $^{56}$Fe we compare the calculated ground-state and finite-temperature cross sections with available results from other approaches.  Conclusions are drawn in Sec.~\ref{conclusion}. In an Appendix we prove in a model independent way that the principle of detailed balance is valid for charge-changing transitions if hot nuclei in the supernova environment are treated in the grand canonical ensemble.

\section{Formalism}\label{formalism}

During the core-collapse phase of a supernovae explosion the temperature in the iron core is sufficiently high (a few $10^9$~K) to establish an equilibrium of reactions mediated by the strong and electro-weak interaction~\cite{Janka_PhysRep442}. Neglecting weak-interaction reactions, we can consider nuclei as open quantum systems in thermal equilibrium with heat and particle reservoirs and, hence, they can be described as a thermal
grand canonical ensemble with temperature $T$ and proton and neutron chemical potentials $\lambda_p$ and $\lambda_n$, respectively. In TFD, such ensemble is represented by the thermal vacuum $|0(T)\rangle$, which is a temperature-dependent state in the extended Hilbert space~\footnote{The correspondence between the thermo-field dynamics and the so-called superoperator formalism is discussed in~\cite{Schmutz_ZPhysB30}. The latter is used by one of the authors (A.D.) to study nonequilibrium transport phenomena (see, e.g.,~\cite{Dzhioev_JPhys24})}.
The thermal vacuum is determined as  the zero-energy eigenstate of the thermal Hamiltonian, ${\cal H}=H-\widetilde H$, and it satisfies the thermal state condition
\begin{equation}\label{TSC}
  A|0(T)\rangle = \sigma_A \mathrm{e}^{\mathcal{H}/2T}\widetilde A^\dag |0(T)\rangle.
\end{equation}
In the above equations $H$ is the original nuclear Hamiltonian (proton and neutron chemical potentials are included in $H$)  and
$\widetilde H$ is its tilde counterpart acting in the auxiliary
Hilbert space; an operator $A$ acts in the physical Hilbert space,
$\widetilde A$ is its tilde partner, and $\sigma_A$ is a phase factor.
The thermal state condition guarantees that the expectation value
$\langle 0(T)|A|0(T)\rangle$ is equal to the  grand canonical
average of $A$. In this sense, relation~\eqref{TSC} is
equivalent to the Kubo-Martin-Schwinger condition for an equilibrium
grand canonical density matrix~\cite{Kubo_JPSJ12,*Martin_PRev115}.

External perturbations mediated by weak interaction induce transitions from the thermal vacuum to nonequilibrium states. Within the TDF such nonequilibrium states are given by the
eigenstates of the thermal Hamiltonian $\cal H$. As follows from the definition
of ${\mathcal H}$, each of its eigenstates with positive energy has
a counterpart --- the tilde-conjugate eigenstate --- with negative but the same absolute value of energy. This gives the possibility to describe both endoergic  and exoergic neutrino reactions with hot nuclei.
In what follows we will refer to positive energy eigenstates as non-tilde states, and to negative energy eigenstates as tilde states.

\subsection{Thermal quasiparticle RPA}\label{TQRPA}

In~\cite{Dzhioev_PAN72, Dzhioev_PRC81} we have introduced the proton-neutron TQRPA method which allows for a treatment of charge-changing transitions in hot nuclei induced by (anti)neutrino absorption. For the sake of completeness, let us briefly recall the method.

Within the TQRPA,  nonequilibrium states of a hot nucleus caused by an external perturbation are treated as
phonon-like excitations on the thermal vacuum
\begin{align}\label{ph_ex}
  &|Q_{JMi}\rangle=Q^\dag_{JMi}|0(T)\rangle,
  \notag\\
  &|\widetilde Q_{JMi}\rangle=\widetilde Q^\dag_{\overline{JM}i}|0(T)\rangle,
\end{align}
where we denote $\widetilde Q^\dag_{\overline{JM}i}=(-1)^{J-M}\widetilde Q^\dag_{J-Mi}$.
Phonon excitations are considered as the normal modes of the thermal Hamiltonian:
\begin{equation}\label{thermal_H_ph}
{\cal H}\simeq\sum_{JM i}\omega_{J i}(T)
   (Q^\dag_{JM i}Q^{\phantom{\dag}}_{JM i}
   -\widetilde Q^\dag_{JM i}\widetilde Q^{\phantom{\dag}}_{JM i}),\end{equation}
while the thermal vacuum $|0(T)\rangle$ itself is the vacuum for the annihilation operators $Q_{JMi}$,  $\widetilde Q_{JMi}$. Thus, within the TQRPA the problem of finding the excitation spectrum of a hot nucleus is reduced to the diagonalization of the thermal Hamiltonian in terms of phonon operators such that the respective phonon vacuum obeys the thermal state condition~\eqref{TSC}.

For charge-changing multipole transitions the phonon operators are defined as a linear superposition
of creation and annihilation operators of proton-neutron thermal quasiparticle pairs~\footnote{In Eq.~\eqref{phonon},
$[\,]^J_M$ denotes the coupling of two single-particle angular
momenta $j_p,\,j_n$ to the total angular~momentum~$J$.}
\begin{multline}\label{phonon}
  Q^\dag_{J M i}=\sum_{j_pj_n}
 \Bigl\{\psi^{Ji}_{j_pj_n}[\beta^\dag_{j_p}\beta^\dag_{j_n}]^J_M +
 \widetilde\psi^{J i}_{j_pj_p}[\widetilde\beta^\dag_{\overline{\jmath_p}}
 \widetilde\beta^\dag_{\overline{\jmath_n}}]^J_M
 \\ +
 i\eta^{J i}_{j_pj_n}[\beta^\dag_{j_p}
  \widetilde\beta^\dag_{\overline{\jmath_n}}]^J_M
+
 i\widetilde \eta^{J i}_{j_pj_n}[\widetilde\beta^\dag_{\overline{\jmath_p}}
\beta^\dag_{j_n}]^J_M
 \\ +
 \phi^{J i}_{j_pj_n}[\beta_{\overline{\jmath_p}}\beta_{\overline{\jmath_n}}]^J_M
+ \widetilde\phi^{J i}_{j_pj_n}[\widetilde\beta_{j_p}
 \widetilde\beta_{j_n}]^J_M \\+
  i\xi^{J i}_{j_pj_n}[\beta_{\overline{\jmath_p}}
  \widetilde\beta_{j_n}]^J_M
 +
  i\widetilde\xi^{J i}_{j_pj_n}[\widetilde\beta_{j_p}
\beta_{\overline{\jmath_n}}]^J_M
  \Bigr\}.
\end{multline}
In turn, thermal quasiparticles are normal modes of the pairing part of the thermal Hamiltonian
\begin{equation}\label{H_pairing}
  \mathcal{H}_\mathrm{pair}\simeq\sum_\tau{\sum_{jm}}^\tau\varepsilon_{j}(T)
  (\beta^\dag_{jm}\beta_{jm} -
  \widetilde\beta^\dag_{j m}\widetilde\beta_{jm}),
\end{equation}
and their vacuum is the thermal vacuum in the BCS approximation. In the expression above, the notation ${\sum}^\tau$ implies a summation over neutron ($\tau=n$) or proton ($\tau=p)$ single particle states only.
Thermal quasiparticles are connected with Bogoliubov quasiparticles by so-called thermal transformation~\footnote{Note that we
use Ojima's\cite{Ojima_AnPhys137} complex form of the thermal transformation.}
\begin{align}
  &\beta^\dag_{jm}=x_j\alpha^\dag_{jm} - i y_j\widetilde\alpha_{jm},
   \notag\\
  &\widetilde\beta^\dag_{jm}=x_j\widetilde\alpha^\dag_{jm} + i y_j\alpha_{jm},~~(x^2_j+y^2_j=1).
\end{align}
The $(x,\,y)$-coefficients as well as the $(u,\,v)$-coefficients of the Bogoliubov transformations are found from the finite-temperature BCS equations~(see~\cite{Dzhioev_PRC81} for more details).
In accordance with the BCS  theory~\cite{Goodman_NPA352,Civitarese_NPA404}, the numerical solution of these equations yields vanishing pairing correlations above a certain critical temperature.

To clarify the physical meaning of different terms in~\eqref{phonon}, we note that the creation of a negative energy tilde thermal quasiparticle corresponds to the annihilation of a thermally excited Bogoliubov quasiparticle or, which is the same, to the creation of a quasihole state (see~\cite{Dzhioev_PRC81} for more details). Therefore, at finite temperature, single-particle charge-changing transitions involve excitations of three types: 1) two-quasiparticle excitations described by the operator $\beta^\dag_{j_p}\beta^\dag_{j_n}$ and having energy $\varepsilon^{(+)}_{j_pj_n}=\varepsilon_{j_p}+\varepsilon_{j_n}$, 2) one-quasiparticle one-quasihole excitations
described by the operators $\beta^\dag_{j_p}\widetilde\beta^\dag_{j_n}$, $\widetilde\beta^\dag_{j_p}\beta^\dag_{j_n}$ and having energies $\varepsilon^{(-)}_{j_pj_n}=\varepsilon_{j_p}-\varepsilon_{j_n}$ and $-\varepsilon^{(-)}_{j_pj_n}$, respectively, and 3) two-quasihole excitations described by the operator $\widetilde\beta^\dag_{j_p}\widetilde\beta^\dag_{j_n}$ and having energy
$-\varepsilon^{(+)}_{j_pj_n}$. The last two types are possible only at $T\ne0$. Therefore, due to single-particle
transitions involving annihilation of thermally excited Bogoliubov quasiparticles, the phonon spectrum at finite temperature contains negative- and low-energy states which do not exist at zero temperature and these "new" phonon states can be interpreted as thermally unblocked transitions between nuclear excited states.

To find the energy and the wavefunctions of thermal phonons we apply  the equation of motion method
\begin{equation}\label{eq_mot}
\langle|\delta Q,[{\cal H},Q^\dag]]|\rangle = \omega(T)\langle|[\delta Q,Q^\dag]|\rangle
\end{equation}
under two additional constraints: (a) the phonon vacuum  obeys
the thermal  state condition~\eqref{TSC}, and (b) phonon operators obey Bose commutation relations.
The first constraint yields the following relations between phonon amplitudes
\begin{align}\label{constraint1}
  \binom{\widetilde\psi}{\widetilde\phi}^{J i}_{j_pj_n}&=\frac{y_{j_p}y_{j_n}-\mathrm{e}^{-\omega_{Ji}/2T}x_{j_p}x_{j_n}}{\mathrm{e}^{-\omega_{Ji}/2T}y_{j_p}y_{j_n}-x_{j_p}x_{j_n}}
\binom{\phi}{\psi}^{J i}_{j_pj_n},
\notag \\
 \binom{\widetilde\eta}{\widetilde\xi}^{J i}_{j_pj_n}&=\frac{y_{j_p}x_{j_n}-\mathrm{e}^{-\omega_{Ji}/2T}x_{j_p}y_{j_n}}{\mathrm{e}^{-\omega_{Ji}/2T}y_{j_p}x_{j_n}-x_{j_p}y_{j_n}}
\binom{\xi}{\eta}^{J i}_{j_pj_n},
\end{align}
while the last assumption is equivalent to averaging with respect to the BCS thermal vacuum in the equations of motion~\eqref{eq_mot} and leads to an orthonormality condition for the amplitudes~\cite{Dzhioev_PRC81}. \
Thus, the phonon energies $\omega$ in~Eq.\eqref{thermal_H_ph} as well as the amplitudes $\psi,\, \widetilde\psi,\,\mathrm{etc.}$ in~ Eq.\eqref{phonon}
are the solution of the proton-neutron TQRPA equations. Positive energy solutions correspond to non-tilde phonons in~\eqref{thermal_H_ph}, while negative energy solutions correspond to tilde ones. Since  both the energies of thermal quasiparticles and the interaction strengths between them are temperature dependent, the spectrum of thermal phonons turns out to be temperature dependent. However, it is significant that in the zero-temperature limit the described method reduces into the standard proton-neutron QRPA.

For a given multipole charge-changing transition operator $\mathcal{T}^{(\mp)}_{J}$ the transition probabilities (strengths) from the thermal vacuum to thermal one-phonon states are given by the following reduced
matrix elements
\begin{align}\label{trans_ampl}
&\Phi^{(\mp)}_{J i}=\bigl|\langle Q_{J
i}\|\mathcal{T}^{(\mp)}_J\|0(T)\rangle\bigr|^2,
  \notag\\
&\widetilde \Phi^{(\mp)}_{J i}=\bigl|\langle\widetilde Q_{J
i}\|\mathcal{T}^{(\mp)}_J\|0(T)\rangle\bigr|^2.
\end{align}
Here the symbol $(-)$ refers to neutron-to-proton transitions ($\beta^-$ channel), while $(+)$ refers to  proton-to-neutron transitions ($\beta^+$ channel). It can be shown that if the transition operators $\mathcal{T}^{(-)}_J$ and $\mathcal{T}^{(+)}_J$ differ only by the isospin operator, i.e., $\mathcal{T}^{(\mp)}_J=\mathcal{T}_Jt_{\mp}$, the transition probabilities to tilde and non-tilde phonon states are connected as
\begin{equation}\label{balance1}
\widetilde\Phi^{(\mp)}_{J i}=\exp\Bigl(-\frac{\omega_{J i}}{T}\Bigr)\Phi^{(\pm)}_{Ji}.
\end{equation}
In~\cite{Dzhioev_PRC89}, a similar relation is obtained for charge-neutral transitions and it is referred to as the principle of detailed balance. However, in contrast to the case of charge-neutral transitions, the detailed balance relation~\eqref{balance1} includes a phonon energy rather than  a transition energy. The latter is the energy transferred to the parent nucleus and for charge-changing transitions is given by~\footnote{At $T=0$, the transition energy corresponds to the final state energy measured from parent-nucleus ground state.}
\begin{align}\label{trans_en}
  E^{(\mp)}_{J i} &= \omega_{Ji}\mp\delta_{np},
  \notag\\
  \widetilde E^{(\pm)}_{J i} & = -E^{(\mp)}_{J i},
  \end{align}
where  $\delta_{np}=\Delta\lambda_{np}+\Delta M_{np}$, and $\Delta\lambda_{np}$ is the difference between  neutron and proton chemical potentials in the parent nucleus and $\Delta M_{np}=1.293$~MeV is the neutron-proton mass splitting. It is obvious that non-tilde (tilde) charge-changing phonon states do not necessarily correspond to upward (downward) transitions: due to $\delta_{np}$, some  non-tilde (tilde) phonon states may have negative (positive) transition energy.
With~Eq.~\eqref{trans_en}, relation~\eqref{balance1} is rewritten as
\begin{equation}\label{balance2}
\widetilde\Phi^{(\mp)}_{J i}=\exp\Bigl(-\frac{E^{(\pm)}_{J i}\mp\delta_{np}}{T}\Bigr)\Phi^{(\pm)}_{Ji}.
\end{equation}
Thus, for each $n\to p$  ($p\to n$) upward transition with energy $E$ there is an inverse downward transition $p\to n$ ($n\to p$) with energy $-E$ and the respective transition probabilities are connected by~\eqref{balance2}.
In the Appendix we show that in this form the detailed balance for charge-changing transitions in the ensemble of hot nuclei can be derived in a model independent way.

\subsection{Cross sections in supernova environments}

In the derivation of the (anti)neutrino absorption cross section for a hot nucleus under supernova conditions
we follow the  Walecka-Donnelly formalism~\cite{Walecka1975,Donnelly_PRep50} which is based on the standard
current-current form of the weak interaction Hamiltonian. After a multipole expansion of the weak nuclear current
the temperature-dependent differential cross section for a transition from the thermal vacuum to a thermal one-phonon state is given by
\begin{align}\label{dif_cr_sect}
  \frac{d\sigma_{Ji}(E_\nu,T)}{d\Omega}&= \frac{(G_F\cos\theta_C)^2}{\pi}\, p^i_e E^i_e \bigl\{\sigma^J_{CL} + \sigma^J_{T}\bigr\}
      \notag\\
      &\times F(\pm Z+1,E_e)(1-f(E_e))\;.
\end{align}
Here and below, the upper (lower) sign refers to the neutrino (antineutrino) cross section. In the above expression, $G_F$ is the Fermi constant for the weak interaction, $\theta_C$ is the Cabbibo angle, and $E_e$ and $p_e$ are the energy and momentum of the  outgoing electron or positron.

In Eq.~\eqref{dif_cr_sect}, the contributions, $\sigma^J_{CL}$, for the Coulomb and longitudinal components, and $\sigma^J_{T}$, for the transverse electric and magnetic components, are written as
\begin{multline}\label{CL}
  \sigma^J_{CL} = (1+a \cos\Theta)|\langle J i\| \hat M_J \|0(T)\rangle|^2+(1+\cos\Theta
  \\
  -2b\sin^2\Theta)|\langle J i\|  \hat L_J\|0(T)\rangle|^2 +\Bigl[\frac{E_{Ji}}{q}(1+\cos\Theta)+c\Bigr]
  \\
  \times2\,\mathrm{Re}\langle J i\| \hat L_J\|0(T)\rangle\langle J i\| \hat M_J\|0(T)\rangle^*,
\end{multline}
\begin{multline}\label{T}
 \sigma^J_{T}=(1-a\cos\Theta+b\sin^2\Theta)\Bigl[|\langle J i\| \hat T^\mathrm{mag}_J\|0(T)\rangle|^2 +\\
   |\langle J i\| \hat T^\mathrm{el}_J\|0(T)\rangle|^2 \Bigr]\mp\Bigl[\frac{E_\nu+E^i_e}{q}(1-a\cos\Theta)-c\Bigr]
  \\
    \times 2\, \mathrm{Re} \langle J i\| \hat T^\mathrm{mag}_J\|0(T)\rangle\langle J i\| \hat T^\mathrm{el}_J\|0(T)\rangle^*,
\end{multline}
where $\Theta$ is the lepton scattering angle and the notation $|Ji\rangle$ is used to denote both the
non-tilde and the tilde thermal one-phonon states. In the former case the transition energy $E_{Ji}=E^{(\mp)}_{Ji}$, while in the latter case $E_{Ji}=\widetilde E^{(\mp)}_{Ji}$. Thus, the energy of the outgoing lepton is $E^i_e=E_\nu-E_{Ji}$. The parameters $a,~b$ and $c$ are obtained from the relations
\begin{align}
  &a=\frac{p_e}{E_e}=\sqrt{1-\Bigl(\frac{m_ec^2}{E_e}\Bigr)^2},
  \notag\\
  &b=a^2\frac{E_\nu E_e}{q^2},~~~~~c=\frac{(m_e c^2)^2}{q E_e},
\end{align}
and the absolute value of the three-momentum transfer $q$ is given
\begin{equation}
 q=\sqrt{E^2_{Ji}+2E_{e}E_\nu(1-a\cos\Theta)-(m_ec^2)^2}\;.
\end{equation}
The multipole operators $\hat M_J$, $\hat L_J$,  $\hat J^\mathrm{el}_J$, and $\hat J^\mathrm{mag}_J$ denote
the charge, longitudinal, and transverse electric and  magnetic parts of the hadronic current, respectively, as definedvin~\cite{Walecka1975,Donnelly_PRep50}. For the vector, axial-vector,  and pseudoscalar form-factors which describebthe internal structure of the nucleon we use parametrization from Ref.~\cite{Singh_NPB77} (see also Ref.~\cite{Djapo_PRevC86}).

For charged-current reactions, the cross section~\eqref{dif_cr_sect} must be corrected for the distortion of the outgoing lepton wave function by the Coulomb field of the residual (daughter) nucleus. The cross section can either be multiplied a Fermi function $F(Z,E)$ (see, e.g., Ref.~\cite{Langanke_NPA481}), or, at higher energies, the effect of the Coulomb field can be described by the effective momentum approximation (EMA)~\cite{Engel_PRC57}. In the present study, by following the prescription from~\cite{Volpe_PRC62,Paar_PRC77}, we choose an energy point in which both approaches predict the same values. Then the Fermi function is used below the point and the EMA is adopted above it.

Furthermore, in the supernova environment nuclei are surrounded by a nearly degenerate electron gas. Thus, neutrino absorption can be strongly reduced by Pauli blocking of the final electron states. The blocking factor  $(1-f(E_e))$ in Eq.~\eqref{dif_cr_sect} accounts for this effects, where $f(E_e)$ is the Fermi-Dirac distribution with temperature $T$ and the chemical potential $\mu_{e^-}$. The positron distribution is defined in the same way with $\mu_{e^+}=-\mu_{e^-}$. Therefore, we can neglect the blocking factor for antineutrino absorption.

The total cross section $\sigma(E_\nu,T)$, as a function of temperature and incoming (anti)neutrino energy, is obtained from the differential cross sections~\eqref{dif_cr_sect} by integrating over the scattering angle and
summing over all possible final thermal one-phonon states
\begin{align}\label{total_CrSect}
  \sigma(E_\nu,T) = &~ 2\pi\sum_{Ji}\int^{-1}_{1} \frac{d\sigma_{Ji}}{d\Omega}\, d\cos\Theta
  \notag\\
  =&\sigma_\mathrm{en}(E_\nu,T) + \sigma_\mathrm{ex}(E_\nu,T).
 \end{align}
Here, the total cross section is split into two parts: $\sigma_\mathrm{en}(E_\nu,T)$ describes the endoergic neutrino absorption and includes only
upward transitions ($E_{Ji}>0$), while  $\sigma_\mathrm{ex}(E_\nu,T)$ corresponds to the exoergic process  associated with
downward transitions ($E_{Ji}<0$).
The contribution   $\sigma_\mathrm{ex}(E_\nu,T)$  dominates the cross section for vanishing neutrino energies, $E_\nu\approx0$. It is apparent that for beta-stable nuclei the exoergic absorption is only possible at finite temperatures and due to transitions from thermally excited states.

For low-energy (anti)neutrinos, i.e., in the long wavelength limit $q\to0$, the allowed $1^+$ multipole operator reduces to the Gamow-Teller form
\begin{equation}\label{GT_op}
  \mathrm{GT}_{\mp}=g_A\vec\sigma t_\mp,
\end{equation}
where $g_A=-1.2599$~\cite{Towner1995} is the  axial weak coupling constant.
Considering only Gamow-Teller transitions and taking into account detailed balance according to~\eqref{balance1}, the cross section~\eqref{total_CrSect} can be written as
\begin{align}\label{GT_cr_sect}
  \sigma&(E_\nu,T)=\frac{(G_F\cos\theta_C)^2}{\pi}
  \notag\\
&\times\Bigl\{{\sum_i} (E_\nu-E^{(\mp)}_i)^2\Phi^{(\mp)}_iF(\pm Z+1,E_e)
\notag\\
&~+ \sum_i (E_\nu+E^{(\pm)}_i)^2\exp\Bigl(-\frac{\omega_i}{T}\Bigr)\Phi^{(\pm)}_iF(\pm Z+1,E_e)\Bigr\}
\end{align}
Here for simplicity the electron rest mass has been neglected, i.e. $E_e\approx p_e$, and the blocking factor for the outgoing lepton has been dropped. The matrix elements $\Phi^{(\mp)}_i$ denote the transition strength of the GT$_\mp$ operator. The first term in Eq.~\eqref{GT_cr_sect} implies summation over non-tilde $1^+$ states. Within the shell model approach this term reduces to the ground-state contribution (see Eq.~(2) in Ref.~\cite{Sampaio_PLett511}). However, in the TQRPA this term appears to be temperature dependent  due to violation of Brink's hypothesis. The second term accounts for transitions to tilde states and it is an analog of the back-resonance contribution within the shell-model approach. Here, we would like to stress that we cannot associate  transitions to non-tilde (tilde)  states with the endoergic (exoergic) absorption.  As was mentioned above, due to $\delta_{np}$ some non-tilde (tilde) states may correspond to downward (upward) transitions and, hence, contribute to $\sigma_\mathrm{ex}$  ($\sigma_\mathrm{en}$).

\section{Results  and discussion}\label{results}

We employ the theoretical framework presented above to study thermal effects on the $(\nu_e,e^-)$ and $(\overline{\nu}_e,e^+)$ cross sections for two sample nuclei, $^{56}\mathrm{Fe}$ and $^{82}\mathrm{Ge}$. The iron isotope is among the most abundant nuclei at the early stage of the core-collapse, while the neutron-rich germanium isotope can be considered as the average nucleus at later stages~\cite{Cooperstein_NPA420}.

To describe charge-changing transitions in a hot nucleus we use the same phenomenological nuclear Hamiltonian
as in~\cite{Dzhioev_PRC81}.  The Hamiltonian consists of spherically symmetric Woods-Saxon potentials for protons and neutrons, BCS pairing interactions, and separable multipole and spin-multipole residuals interaction in the particle-hole channel. We neglect particle-particle interactions except for the BCS pairing forces. This Hamiltonian is usually referred to as the quasiparticle-phonon model (QPM)~\cite{Soloviev1992}. For the two nuclei considered, the parameters of the QPM Hamiltonian are fixed in the same manner as
in~\cite{Dzhioev_PRC81,Dzhioev_PRC89}. Here we just mention that solving the BCS equations at zero temperature we get the following proton and neutron pairing gaps: $\Delta_{p(n)} = 1.57(1.36)$~MeV for
$^{56}$Fe and $\Delta_{p(n)}= 1.22(0.0)$~MeV for $^{82}$Ge. Thus, the critical temperature ($T_\mathrm{cr}\approx0.5\Delta_\tau$) when the pairing phase transition occurs is $T_\mathrm{cr}\approx0.8$~MeV for the iron isotope and $T_\mathrm{cr}\approx0.6$~MeV for the germanium isotope.

A separable form of the residual interaction allows us to reduce the TQRPA equations to a secular equation. The explicit form for charge-changing excitations as well as expressions for transition strengths~\eqref{trans_ampl}
are given in~\cite{Dzhioev_PAN72,Dzhioev_PRC81}. Moreover, it has been proved in Ref.~\cite{Dzhioev_PAN72} that the TQRPA fulfills the Ikeda sum rule for allowed Fermi and GT transitions.

\subsection{Ground-state cross sections}

\begin{figure*}[!t]
 \begin{centering}
\includegraphics[width=0.7\textwidth]{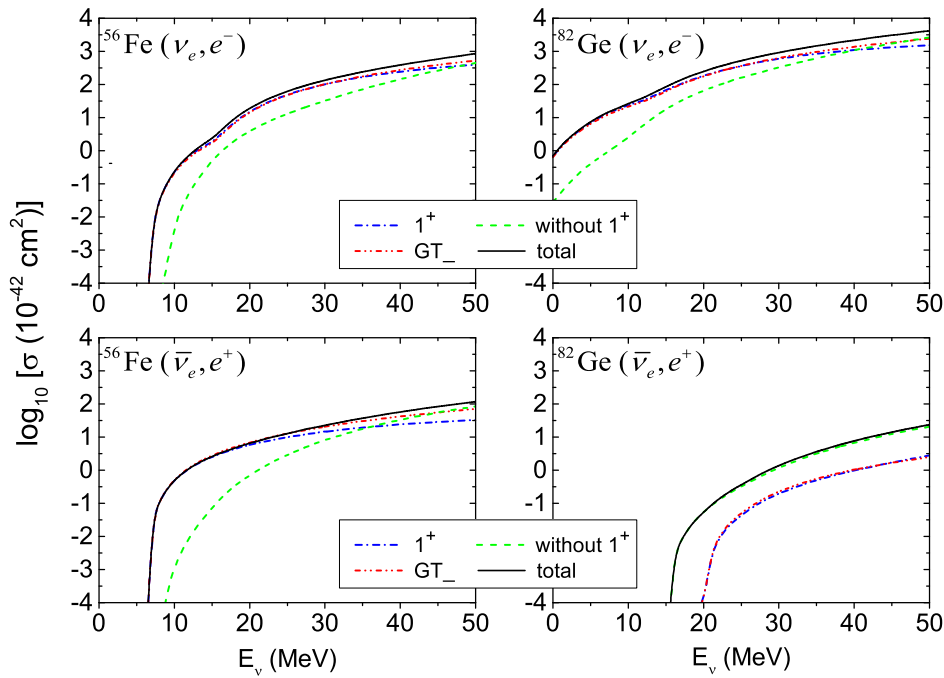}
\caption{(Color online) Cross sections for neutrino (upper panels) and antineutrino (lower panels) absorption reactions for $^{56}$Fe and $^{82}$Ge. The total
cross sections (solid lines)  include contributions of the $J^\pi=0^\pm-3^\pm$ multipoles. The dashed lines show the cross sections without the $1^+$ contribution.
The dash-dotted lines correspond to the $1^+$ contributions calculated with the full momentum-dependent transition operator whereas the $1^+$ contributions
calculated with the Gamow-Teller operator~\eqref{GT_op} are shown by the dash-double dotted lines.}
 \label{GS_CrSect}
 \end{centering}
\end{figure*}

As in~\cite{Dzhioev_PRC89}, before proceeding to discuss thermal effects on neutrino-nucleus absorption cross sections, we consider the ground-state ($T=0$) cross sections for $^{56}\mathrm{Fe}(\nu_e,e^-)$,
$^{56}\mathrm{Fe}(\overline{\nu}_e,e^+)$, $^{82}\mathrm{Ge}(\nu_e,e^-)$, and
$^{82}\mathrm{Ge}(\overline{\nu}_e,e^+)$ reactions and compare the our results with those available from other theoretical studies. In Fig.~\ref{GS_CrSect}, the calculated total cross sections are shown as functions of the incident (anti)neutrino energy $E_\nu$. We find that for $E_\nu\le 100\,\mathrm{MeV}$  the reactions considered are dominated by the multipole transitions $J^\pi\le 3^\mp$, while contributions from higher multipoles are only
a few percent of the total cross sections.

For  $^{56}$Fe the cross sections increase sharply as  $E_\nu$ approaches the reaction threshold $Q_-=4.56$~MeV for $\nu_e$-absorption and $Q_+=4.71$~MeV for $\overline{\nu}_e$-absorption ($Q_\mp= M_f+m_ec^2-M_i$, where $M_{i,f}$ are the masses of the parent and daughter nuclei). As the neutron number increases, the threshold energy for $\nu_e$-absorption decreases. For the neutron-rich nucleus $^{82}\mathrm{Ge}$ the $Q_-$-value becomes negative ($Q_-=-4.71$~MeV) allowing neutrino absorption for all energies. Contrary to this, for the $(\overline{\nu}_e,e^+)$ reactions the $Q_+$-values become less favorable with increasing neutron excess. The antineutrino has to overcome a noticeable threshold ($Q_+=13.58$~MeV) to be absorbed  by $^{82}\mathrm{Ge}$ and the corresponding cross section is considerably lower in comparison to the neutrino one.

In addition to the total cross sections, contributions from the $1^+$ multipole channel are shown in Fig.~\ref{GS_CrSect}. Referring to the plots in Fig.~\ref{GS_CrSect} for the  $^{56}\mathrm{Fe}(\nu_e,e^-)$,
$^{56}\mathrm{Fe}(\overline{\nu}_e,e^+)$, $^{82}\mathrm{Ge}(\nu_e,e^-)$ reactions the cross sections are dominated by allowed $1^+$  transitions for energies up  to $E_\nu = 30$~MeV,  with contributions from other multipoles being  much smaller. In the $^{82}\mathrm{Ge}(\overline{\nu}_e,e^+)$ cross section, however, the $1^+$ contribution is negligible and forbidden transitions dominate the reaction. We find that $1^-$ and $2^-$ transitions mainly contribute to the cross section for $E_\nu < 50$~MeV.

In Fig.~\ref{GS_CrSect} we also analyze the effect of the full $q$-dependent $1^+$ transition operator instead of its long-wavelength limit. As was mentioned above, in the latter case the $1^+$ transition operator reduces to the Gamow-Teller operator~\eqref{GT_op} and the ground-state cross section is given by the first term in Eq.~\eqref{GT_cr_sect}. It should be noted that to make a comparison with the shell-model calculations~\cite{Kolbe_PRC63,Toivanen_NPA694} more transparent, we use the same quenching factor for the axial weak coupling constant, $g_A^*=0.74g_A$. By comparing in Fig.~\ref{GS_CrSect} the $1^+$ and Gamow-Teller contributions to the cross sections, we conclude that for energies $E_\nu\le 30$~MeV the application of the GT operators instead of the $q$-dependent $1^+$ operator is fully justified. Therefore, we conclude that
the low-energy ground-state cross sections for the $^{56}\mathrm{Fe}(\nu_e,e^-)$,
$^{56}\mathrm{Fe}(\overline{\nu}_e,e^+)$, $^{82}\mathrm{Ge}(\nu_e,e^-)$ reactions are completely dominated  by GT transitions. In contrast, all GT$_+$  transitions of are essentially blocked in $^{82}$Ge. For nuclei with $Z<40$ and $N>40$ blocking occurs because the valence protons are in the $pf$ shell, while the valence neutrons occupy already the next major shell ($sdg$ shell). In the next section it will be shown, however, that thermal effects unblock GT$_+$ transitions in $^{82}$Ge and for typical supernova temperatures the low-energy $^{82}\mathrm{Ge}(\overline{\nu}_e,e^+)$ cross sections are also dominated by the GT contributions.

\begin{table}[t]
    \caption{Total $^{56}\mathrm{Fe}(\nu_e,e^-)$ cross sections for selected
    neutrino energies $E_\nu$.
    The present QRPA results (second column) are compared with
    those from~\cite{Lazauskas_NPA792}
     and with the hybrid approach results~\cite{Kolbe_PRC63}. The cross sections are given in units of $10^{-42}~\mathrm{cm}^2$, with exponents given in parentheses.}
  \begin{tabular}{cccc}
    \hline\hline
    ~$E_\nu$ (MeV)~&~QRPA~&~SkQRPA\cite{Lazauskas_NPA792}~&~Hybrid\cite{Kolbe_PRC63}~ \\
    \hline\hline
    10 & 2.39($-$1)  & 3.63(+0)  & 6.61($-$1)   \\
    15 & 2.35(+0)  & 1.73(+1)  & 6.45(+0)   \\
    20 & 1.91(+1)  & 5.26(+1)  & 2.93(+1)   \\
    25 & 6.19(+1)  & 1.25(+2)  & 7.33(+1)   \\
    30 & 1.34(+2)  & 2.33(+2)  & 1.40(+2)   \\
    40 & 3.88(+2)  & 5.44(+2)  & 3.71(+2)   \\
    50 & 8.47(+2)  & 9.83(+2)  & 7.98(+2)   \\
    60 & 1.58(+3)  & 1.67(+3)  & 1.38(+3)   \\
    70 & 2.61(+3)  & 2.59(+3)  & 2.42(+3)   \\
    80 & 3.95(+3)  & 3.73(+3)  & 3.60(+3)   \\
    90 & 5.53(+3)  & 5.07(+3)  & 4.98(+3)   \\
   100 & 7.26(+3)  & 6.60(+3)  & 6.52(+3)   \\
   \hline\hline
  \end{tabular}
 \label{56Fe_neutrino}
\end{table}

\begin{table}[t]
    \caption{Same as Table~\ref{56Fe_neutrino}, but for the $^{56}\mathrm{Fe}(\overline{\nu}_e,e^+)$ reaction.}
  \begin{tabular}{cccc}
    \hline\hline
    ~$E_\nu$ (MeV)~&~QRPA~&~SkQRPA\cite{Lazauskas_NPA792}~&~Hybrid\cite{Toivanen_NPA694}~ \\
    \hline\hline
    10.0    & 4.92($-1$)  & 2.95(+0)  & 0(+0)   \\
    12.5    & 1.34 (+0)   & 6.09(+0)  & 0(+0)   \\
    15.0    & 2.59(+0)    & 1.03(+1)  & 2(+0)   \\
    20.0    & 6.48(+0)    & 2.17(+1)  & 7(+0)   \\
    25.0    & 1.28(+1)    & 3.74(+1)  & 1.6(+1) \\
    30.0    & 2.26(+1)    & 5.74(+1)  & 3.0(+1)  \\
    40.0    & 5.71(+1)    & 1.20(+2)  & 8.4(+1)   \\
    50.0    & 1.17(+2)    & 2.09(+2)  & 1.81(+2)   \\
   \hline\hline
  \end{tabular}
 \label{56Fe_antineutrino}
\end{table}

In Tables~\ref{56Fe_neutrino} and~\ref{56Fe_antineutrino}, we compare the calculated ground-state  cross sections for (anti)neutrino absorption by $^{56}$Fe to those obtained from self-consistent QRPA
calculations with Skyrme forces~\cite{Lazauskas_NPA792} and with the hybrid approach (large-scale shell-model calculations for GT contributions and RPA for other multipoles)~\cite{Kolbe_PRC63,Toivanen_NPA694}. As is seen, the three models, although based on different microscopic pictures,
predict rather similar neutrino absorption cross sections. We note, however, that for low energies ($E_\nu\le 40$~MeV) the present results are closer to the hybrid approach results than those of Ref.~\cite{Lazauskas_NPA792}. It is evident that the reason why the  QRPA and the hybrid approach cross sections  are systematically lower than those of the Skyrme based calculations
at $E_\nu\le 40$~MeV is caused by differences in the GT$_-$ strength distributions.
Although the authors of Ref.~\cite{Lazauskas_NPA792} do not provide the GT$_-$  distribution in $^{56}$Fe,
it seems that the Skyrme based calculations  result in a larger strength below the GT$_-$ resonance compared to the other two approaches.
Other possible reasons for the discrepancy could be a lower energy of the GT$_-$ resonance and a larger total GT strength obtained from the calculations with the Skyrme interaction.

Experimental results for neutrino absorption cross sections are rather limited. The KARMEN Collaboration has measured the flux averaged $^{56}\mathrm{Fe}(\nu_e, e^-)^{56}\mathrm{Co}$ cross section for the neutrino spectrum from the muon decay at rest
and obtains $\langle\sigma\rangle = [251\pm83\mathrm{(stat.)}\pm42\mathrm{(syst.)}]\times10^{-42}\,\mathrm{cm}^2$\cite{Maschuw_PPNP40}. Our result, $\langle\sigma\rangle = 223\times10^{-42}\mathrm{cm}^2$, is in excellent agreement with the experimental value. Note, that
the hybrid approach predicts $\langle\sigma\rangle = 240\times10^{-42}\mathrm{cm}^2$~\cite{Kolbe_PRC63}, while the QRPA calculations with Skyrme forces gives $\langle\sigma\rangle = 352\times10^{-42}\mathrm{cm}^2$~\cite{Lazauskas_NPA792}.

\begin{figure}[t]
 \begin{centering}
\includegraphics[width=1.0\columnwidth]{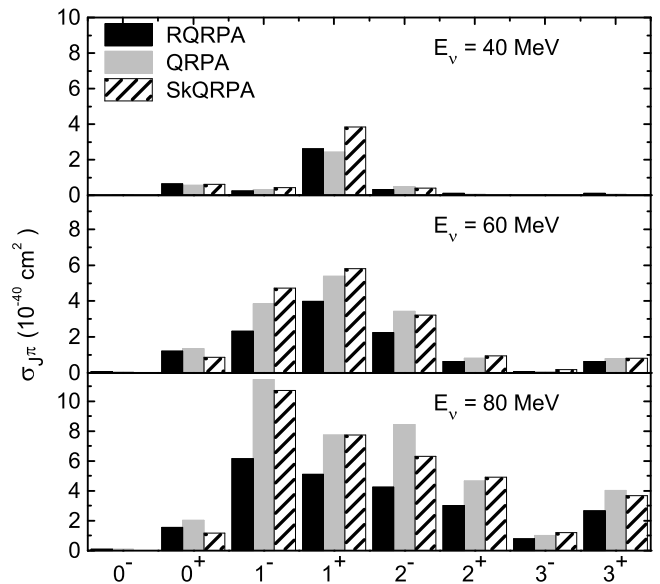}
\caption{Contribution of the multipole transitions $J^\pi=0^\pm-3^\pm$ to the cross section for  the $^{56}\mathrm{Fe}(\nu_e,e^-)^{56}\mathrm{Co}$
reaction at $E_\nu=40,~60,~80$~MeV. The present QRPA results are compared with those obtained from RQRPA~\cite{Paar_PRC84}
and SkQRPA~\cite{Lazauskas_NPA792} calculations. }
 \label{partial}
 \end{centering}
\end{figure}

For  $^{56}$Fe a detailed demonstration of the most important multipole contributions to neutrino absorption cross section is presented in Fig.~\ref{partial} at three neutrino energies, $E_\nu=40,~60,\mathrm{and}~80$~MeV.
As expected, at relatively low neutrino energies ($E_\nu\lesssim 40$~MeV) the dominant contribution to the cross section originates from $1^+$ transitions. With increasing $E_\nu$, however, contributions from other multipole transitions become important. In particular, at $E_\nu=80$~MeV, the dominant contribution comes from $1^-$ transitions, but other multipoles, e.g.,  $1^+$, $2^-$, $2^+$, and $3^+$, also play an important role.

In Fig.~\ref{partial}, we also compare the calculated multipole decomposition of the cross section for $^{56}$Fe with those from the SkQRPA calculations~\cite{Lazauskas_NPA792} and the relativistic QRPA (RQRPA) calculations~\cite{Paar_PRC84}.  As is evident from the figure, the latter model predicts somewhat smaller cross section at high neutrino energies ($E_\nu\gtrsim60$~MeV), whereas we observe an excellent agreement between the results of the present QRPA and the SkQRPA. Specifically, in accordance with Ref.~\cite{Lazauskas_NPA792, Paar_PRC84}, we find that $0^+$ allowed transitions only marginally contribute to the
$^{56}\mathrm{Fe}(\nu_e,e^-)$ reaction. This finding is true for other three reactions and for finite temperatures as well. For this reason, in the discussion below, we will always imply $1^+$ multipole channels when referring the allowed transitions.

\subsection{Finite temperatures}

In discussing $\nu_e$ and $\overline{\nu}_e$ absorption reactions under supernova conditions we will follow the line of our recent work~\cite{Dzhioev_PRC89} and consider first thermal effects on the strength distribution of GT transitions which dominate the reactions at low energies.

\begin{figure}[t!]
 \begin{centering}
\includegraphics[width=\columnwidth, angle=0]{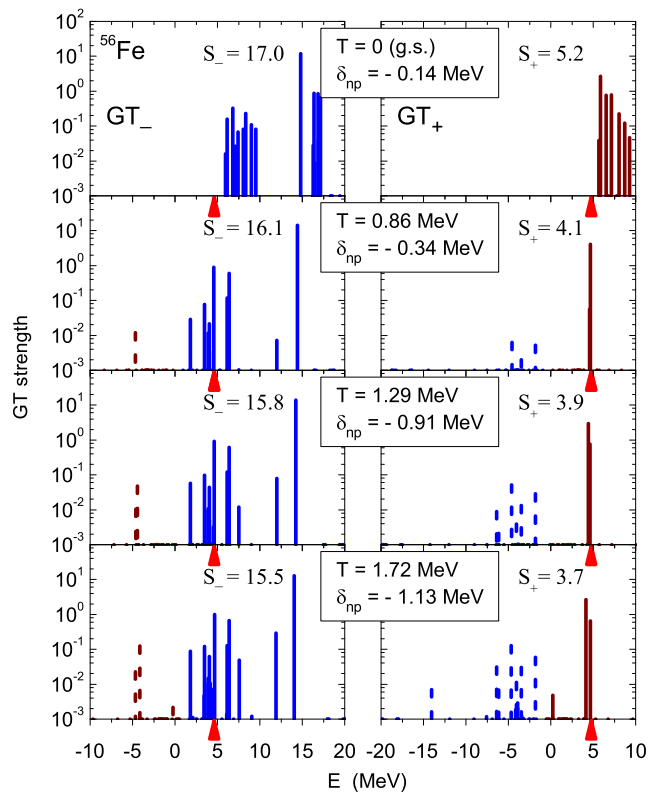}
\caption{(Color online) Temperature evolution of GT$_-$ (left panels) and GT$_+$ (right panels) strength distributions for $^{56}$Fe vs transition energy. The solid (dashed) lines refer to transitions to non-tilde (tilde) thermal one-phonon states. The arrows indicate the ground-state reaction thresholds for neutrino ($Q_-=4.56$~MeV) and  antineutrino ($Q_+=4.71$~MeV) absorption. }
\label{GT_56Fe_Temp}
 \end{centering}
\end{figure}

In Fig.~\ref{GT_56Fe_Temp}, we display on a logarithmic scale the GT$_-$ and GT$_+$ distributions in $^{56}$Fe calculated for the ground-state ($T=0$) and at
three stellar temperatures above the critical one, $T=0.86\,\mathrm{MeV}\,(10^{10}\,\mathrm{K})$, $1.29\,\mathrm{MeV}\,(1.5\times10^{10}\,\mathrm{K})$, and $1.72\,\mathrm{MeV}\,(2\times10^{10}\,\mathrm{K})$.
These temperatures  roughly correspond to three stages in the collapse evolution~\cite{Juodagalvis_NPA747}. We emphasize that the distributions are plotted with the bare GT operators $\vec\sigma t_\mp$ as functions of the transition energy~\eqref{trans_en}. For each temperature we show the value of $\delta_{np}$ in Eq.~\eqref{trans_en} as well as the total transition strengths $S_-$ and $S_+$. Note that $S_-$ and $S_+$ values calculated for the ground-state  satisfy the Ikeda sum rule $S_- - S_+=3(N-Z)$ (a small deviation is caused by the incompleteness of our single-particle basis) but noticeably overestimate the experimental data ($S_-=9.9\pm2.4$~\cite{Rapaport_NPA410}, $S_+=2.9\pm0.3$~\cite{ElKateb_PRC49}). This overestimation is common for any QRPA calculations of GT strength and is remedied by a quenched value for the axial weak coupling constant $g_A$.
One might notice that our QRPA calculations fairly well reproduce the experimental centroid energies for both GT$_+$~\cite{ElKateb_PRC49} and  GT$_-$~\cite{Rapaport_NPA410} distributions in $^{56}$Fe.  In this respect the present calculations are consistent with the large-scale shell-model calculations~\cite{Caurier_NPA653}. Of course, in contrast to the LSSM approach, the QRPA cannot recover all nuclear correlations needed to reproduce the fragmentation of the GT strength.

For $^{56}$Fe, due to a relatively small absolute value of $\delta_{np}$, the major part of the upward (downward) GT strength  corresponds to transitions to non-tilde (tilde) phonon states. According to our QRPA calculations,  the principal contribution to the GT$_-$  (GT$_+$) strength in $^{56}$Fe comes from the $1f_{7/2}\to 1f_{5/2}$ neutron-to-proton (proton-to-neutron) single-particle transition which forms a resonance peak at~$E\approx 14$~MeV ($\approx 6$~MeV). Because the Brink hypothesis is not valid within the TQRPA, we observe an evolution of the GT resonances with temperature. Namely, when the temperature is increased to $0.86$~MeV, the transition energy is lowered by $\sim1.2$~MeV for the GT$_-$ resonance and $\sim2.1$~MeV for the GT$_+$ resonance. This decrease in energy is mainly attributed to the vanishing of pairing correlations:  at temperatures above the critical one no extra energy is needed to break a proton (neutron) Cooper pair and as a consequence the GT$_+$ (GT$_-$) resonance moves to lower energies. However, not only pairing effects lead to downward shift of the GT resonances. It can be easily seen from the structure of the TQRPA equations~\cite{Dzhioev_PAN72} that quasiparticle thermal occupations factors will appear which screen the interaction term. Due to the  thermal blocking of the proton-neutron repulsive residual interaction, a further increase in temperature to $1.72$~MeV decreases the GT$_-$ and GT$_+$ resonances in $^{56}$Fe by $\sim 0.3$~MeV and $\sim 0.5$~MeV, respectively. As was mentioned in the introduction, the observed downward shift of the GT strength is not present in large-scale shell-model calculations which are partially based on Brink's hypothesis. In contrast, the finite-temperature relativistic QRPA calculations in Ref.~\cite{Niu_PRC83} and shell-model Monte-Carlo calculations in Ref.~\cite{Radha_PRC56} show similar features for the changes of the GT resonance energy.

Finite temperature also affects the low-energy GT$_-$ strength in $^{56}$Fe: due to the gradual disappearance of pairing it partially shifts below the ground-state reaction threshold and becomes more fragmented. The fragmentation arises from the thermal smearing of the nuclear Fermi surface that unblocks low-energy particle-particle and hole-hole GT$_-$ transitions. Here, particle (hole) denotes a state above (below) the Fermi level. Moreover, in accordance with detailed balance~\eqref{balance2}, the temperature rise exponentially increases the strength of negative-energy downward transitions. Referring to the figure, the dominant part of the downward GT$_-$ strength around $E\approx -4.1$~MeV originates from a transition inverse to the GT$_+$ resonance, that is due to the transition from the thermally populated neutron orbit $1f_{5/2}$  to the lower lying proton orbit  $1f_{7/2}$. In contrast, transitions inverse to the GT$_-$ resonance are suppressed by the energy-dependent exponent in Eq.~\eqref{balance2} and the GT$_+$ downward strength is dominated by transitions inverse to low-energy GT$_-$ transitions. It should be mentioned that although thermal effects unblock some new GT transitions, the total GT$_\mp$ strength in $^{56}$Fe slightly decreases with temperature. Nevertheless,  the TQRPA preserves the Ikeda sum rule.

It is obvious that the violation of Brink's hypothesis should affect the downward GT strength. In~\cite{Dzhioev_PRC89}, this influence was studied thoroughly for charge-neutral GT transitions in $^{56}$Fe by comparing the GT running sums derived with and without Brink's hypothesis. Applying the same method, we find that both the shift of the GT$_+$ resonance  to lower energies and the thermal unblocking of low-lying GT$_-$ strength enhance the GT$_-$ and GT$_+$ downward strengths in $^{56}$Fe.

\begin{figure}[t!]
 \begin{centering}
\includegraphics[width=\columnwidth, angle=0]{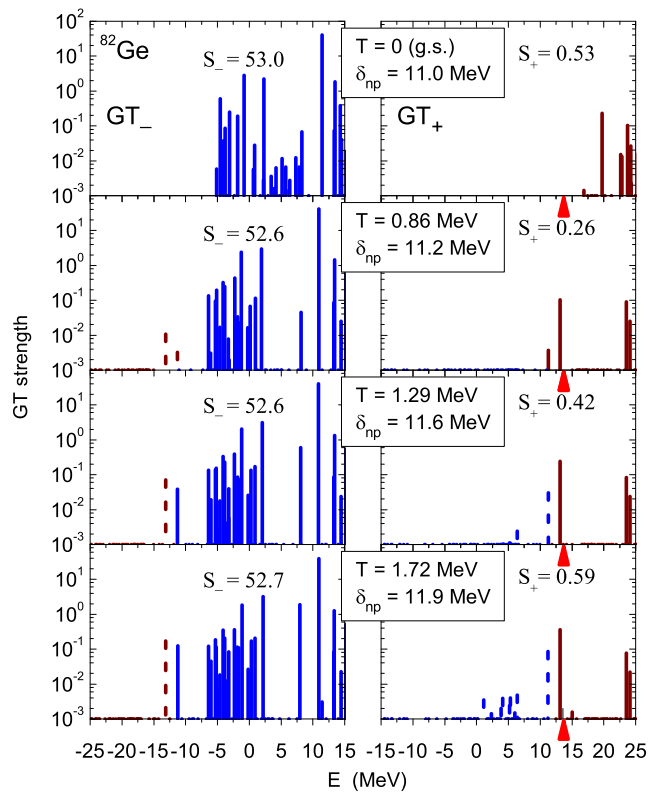}
\caption{(Color online) Same as Fig.~\ref{GT_56Fe_Temp} but for $^{82}$Ge.
Due to Pauli blocking of allowed $p\to n$ transitions, the total GT$_-$ strength is many times larger than the GT$_+$ strength and $S_-\approx 3(N-Z)=54$. } \label{GT_82Ge_Temp}
 \end{centering}
\end{figure}

Fig.~\ref{GT_82Ge_Temp} shows the temperature evolution of the GT strength function for the neutron-rich nucleus $^{82}$Ge. The distributions are displayed at the same temperatures as those for $^{56}$Fe  in Fig.~\ref{GT_56Fe_Temp}. As evident from the figure, some GT strength associated with non-tilde (tilde) phonon states appears at negative (positive) transition energies which is due to a large value of $\delta_{np}$. The temperature rise affects the GT$_-$ and GT$_+$ distributions in a different way. For the GT$_-$ distribution, a temperature increase essentially affects only the downward part, yielding some new strength below $-10$~MeV. By comparing the left and right panels in Fig.~\ref{GT_82Ge_Temp}, we conclude that this thermally unblocked strength corresponds to  transitions inverse to the GT$_+$ resonance. The GT$_-$ resonance is weakly sensitive to thermal effects. By increasing the temperature to $T=1.72$~MeV, the excitation energy gets only lowered by $\sim 0.5$~MeV. Since there are no neutron pairing correlations in $^{82}$Ge, this lowering is  solely caused by the softening of the residual interaction.

In contrast,  thermal effect are significant for the GT$_+$ resonance and they clearly demonstrate a violation of the Brink hypothesis within the TQRPA. Namely, with rising temperature the resonance peak shifts by $\sim 7$~MeV to lower energies and the total  GT$_+$ strength shows a nonmonotomic dependence: after an initial decrease, it gradually increases. In~\cite{Dzhioev_PRC81}, a detailed investigation within the TQRPA approach was performed of thermal effects on the GT$_+$ strength in the neutron-rich nucleus $^{76}$Ge. The conclusions  remain valid for  $^{82}$Ge, as well. Briefly, for neutron-rich nuclei with $N>40$ and $Z<40$ the Independent Particle Model predicts that all GT$_+$ transitions of valence protons are Pauli blocked due to the complete occupation of the $pf$ neutron orbits. However, both thermal excitations and pairing correlations promote protons to the $sdg$ shell and make possible $1g^p_{9/2}\to1g^n_{7/2}$ particle-particle transition in $^{82}$Ge possible. The important point is that the transition energy drastically decreases with temperature. Namely, at $T<T_\mathrm{cr}$, when pairing correlations are important, the transition energy is given by $\varepsilon_{1g^n_{7/2}}+\varepsilon_{1g^p_{9/2}}+\delta_{pn}\approx 20$~MeV, while at  $T>T_\mathrm{cr}$, when thermal effects become significant, it is given by $\varepsilon_{1g^n_{7/2}}-\varepsilon_{1g^p_{9/2}}+\delta_{pn}\approx 13$~MeV.
In addition, the transition strength below the GT$_+$ resonance becomes increasingly unblocked with temperature. In particular, a strong peak due to  the $1f^p_{7/2}\to1f^n_{5/2}$ hole-hole transition appears at $E\approx11$~MeV. Both unblocking mechanisms are severely suppressed in the vicinity of the critical temperature
($T_\mathrm{cr}\approx0.6$~MeV in $^{82}$Ge), i.e. when pairing correlations vanish while thermal effects are not yet sufficiently strong to significantly occupy the $1g_{9/2}$ proton orbit or unblock the $1f_{5/2}$ neutron orbit. Therefore, the total  GT$_+$ strength decreases at $T\approx T_\mathrm{cr}$. The thermal effects on the GT$_+$ strength in neutron-rich nuclei  discussed here were predicted in~\cite{Cooperstein_NPA420} and also confirmed in~\cite{Niu_PRC83} by calculations based on the finite-temperature relativistic QRPA. We also note that, due to the large $\delta_{np}$, downward GT$_+$ transitions appears to be highly suppressed in $^{82}$Ge.

\begin{figure}[t!]
 \begin{centering}
\includegraphics[width=\columnwidth, angle=0]{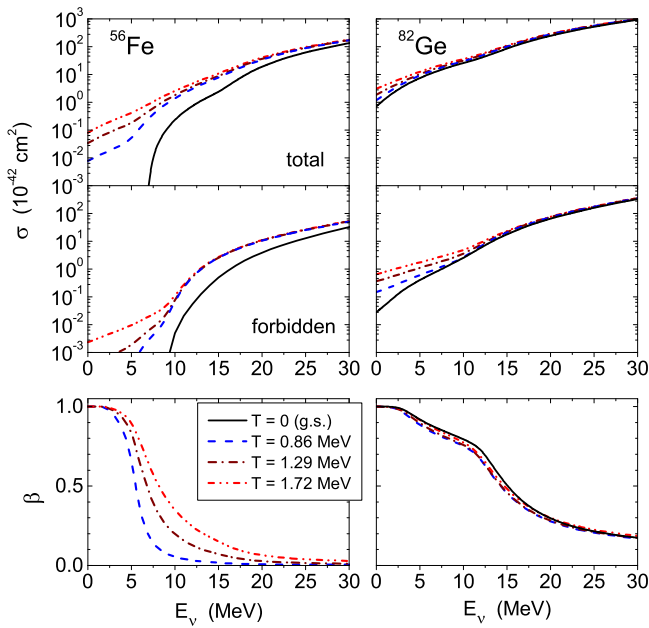}
\caption{(Color online) (Upper panels) Total neutrino absorption cross sections for $^{56}$Fe and $^{82}$Ge at three different temperatures relevant for core collapse. For comparison, the ground-state cross sections are also shown. (Middle panels) Contributions from forbidden transitions to the cross sections. (Lower panels) Temperature dependence of  the ratio of exoergic absorption to the reaction cross section.}
\label{nnb}
 \end{centering}
\end{figure}

Let us now illustrate the influence of the thermal effects discussed above on the neutrino and antineutrino absorption by  $^{56}\mathrm{Fe}$ and $^{82}\mathrm{Ge}$. In Fig.~\ref{nnb}, we compare the $(\nu_e, e^-)$ ground-state cross sections with those calculated at the three core-collapse temperatures. For energies $E_\nu<30~\mathrm{MeV}$, which are typical for supernova neutrinos, the cross sections are dominated by allowed  GT$_-$ transitions at all temperatures. This is verified in the middle panels of  Fig.~\ref{nnb}, where the overall contribution of forbidden transitions to the finite temperature cross sections is given. Although the forbidden cross section increases with temperature, it contributes less than 30\% all the way up to $E_\nu=30~\mathrm{MeV}$.
The lower panels of Fig.~\ref{nnb} show the ratio of exoergic absorption to the  reaction cross section
\begin{equation}
  \beta(E_\nu,T) = \frac{\sigma_\mathrm{ex}(E_\nu,T)}{\sigma(E_\nu,T)},
\end{equation}
where $\sigma_\mathrm{ex}(E_\nu,T)$ only accounts for negative-energy downward transitions induced by neutrino absorption.

When considering the $^{56}\mathrm{Fe}(\nu_e, e^-)$ cross section,  we observe that thermal effects are unimportant for $E_\nu > 20$~MeV, i.e. when incoming neutrinos have sufficiently large energy to excite the GT$_-$ resonance. Note, that a downward shift of the GT$_-$ resonance only marginally affects the cross section at such high neutrino energies. Thermal effects become pronounced though for lower neutrino energies since finite temperature removes the reaction threshold and significantly enhances the cross section. For energies $E_\nu<5$~MeV, the ratio $\beta>0.5$ and, hence, the observed enhancement is mostly caused by downward GT$_-$ transitions from nuclear excited states. Moreover, for $E_\nu\approx 0$ the thermally unblocked downward transitions completely dominate the reaction, increasing  the cross section by more than an order of magnitude when the temperature rises from 0.86~MeV to 1.72~MeV. The role of the exoergic absorption decreases with increasing neutrino energy and for intermediate energies, $5\,\mathrm{MeV}<E_\nu<20\,\mathrm{MeV}$,
thermal effects on the GT$_-$  resonance and its low-energy tail become important for the cross section
enhancement.

Referring to the right panels in Fig.~\ref{nnb} it is shown that the $^{82}\mathrm{Ge}(\nu_e, e^-)$ cross section to a much lesser extent depends on temperature than that for the $^{56}\mathrm{Fe}(\nu_e, e^-)$ reaction. This result can be understood as follows. In $^{82}\mathrm{Ge}$, due to the negative $Q_-$-value, the downward GT$_-$ transitions dominate the ground-state reaction up to $E_\nu\approx 15$~MeV. As discussed above, the GT$_-$ distribution in  $^{82}\mathrm{Ge}$ is little affected by the temperature rise, which merely causes some additional strength at $E<-10$~MeV. This thermally unblocked downward strength becomes competitive with the ground-state contribution only at rather low energies ($E_\nu\lesssim 5$~MeV). As a result, thermal effects on the $^{82}\mathrm{Ge}(\nu_e, e^-)$ reaction are noticeably milder than in the previous case and a temperature rise from $T=0$ to 1.72~MeV enhances the low-energy cross section only by a factor of about four. This observation is in line with Ref.~\cite{Sampaio_PLett511}, where it was shown that the thermal enhancement of the neutrino absorption cross section is reduced the more neutron-rich the nucleus is.

\begin{figure}[t!]
 \begin{centering}
\includegraphics[width=\columnwidth, angle=0]{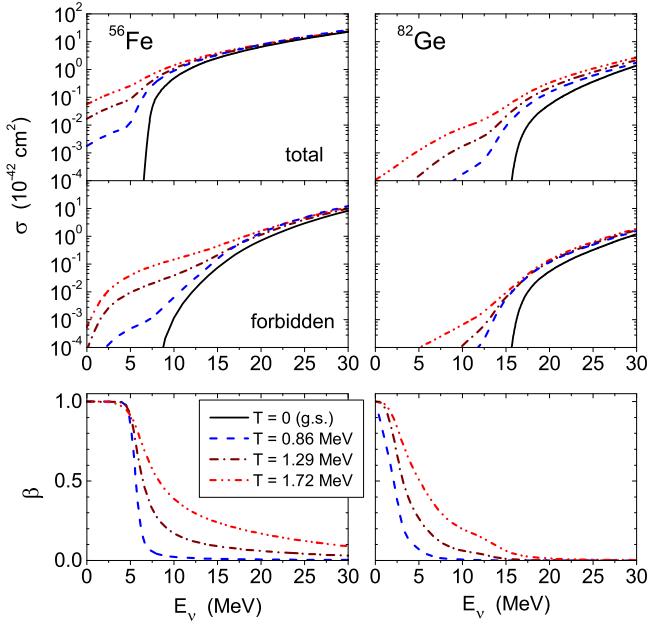}
\caption{(Color online) The same as in Fig.~\ref{nnb}, but for the antineutrino absorption reaction.} \label{annb}
 \end{centering}
\end{figure}

The results for $\overline{\nu}_e$-absorption are shown in Fig.~\ref{annb}. For the $^{56}\mathrm{Fe}(\overline{\nu}_e, e^+)$ reaction the cross section demonstrates the same trend as discussed above for the neutrino absorption by $^{56}\mathrm{Fe}$: (i) The gap in the cross section disappears and the low-energy cross section increases with temperature. (ii) with increasing $E_\nu$  the cross sections at different temperatures converge, i.e., thermal effects become unimportant. From the lower-left panel we conclude that a significant enhancement of the low-energy cross section relative to ground-state calculations comes from the increasing contribution of downward GT$_-$ transitions from nuclear excited states. The ratio $\beta$  gradually reduces  at $E_\nu>5$~MeV, since at those energies the excitation of the GT$_+$ resonance becomes possible. However, at $T=1.72$~MeV the exoergic component of the cross section appears to be comparable with the endoergic one up to $E_\nu\approx15$~MeV.

\begin{figure}[t!]
 \begin{centering}
\includegraphics[width=\columnwidth, angle=0]{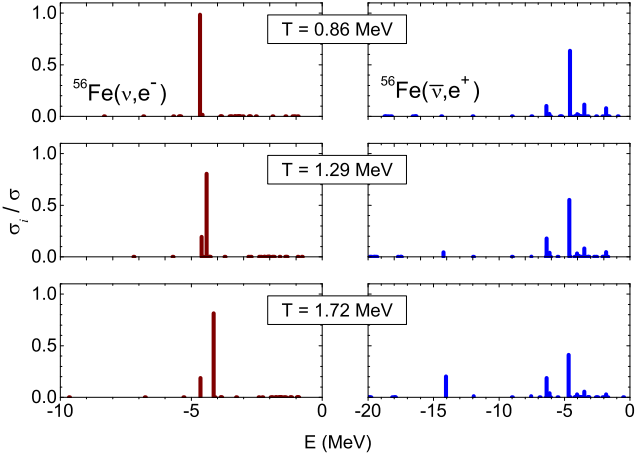}
\caption{(Color online) Relative contribution $\sigma_i/\sigma$ of the $i$th negative-energy GT state to the finite-temperature cross section at $E_\nu=0$.} \label{partial_T}
 \end{centering}
\end{figure}

It should be noted that although the calculated $^{56}\mathrm{Fe}({\nu}_e, e^-)$ and $^{56}\mathrm{Fe}(\overline{\nu}_e, e^+)$ cross sections show the same trend, the observed thermal enhancement at low energies  is caused by different types of downward transitions. To show that, we have calculated the relative contribution $\sigma_i/\sigma$ of different negative-energy GT states to the cross section at $E_\nu=0$. The results are depicted in Fig.~\ref{partial_T}. As indicated in the figure, the $\nu$-absorption is completely dominated by the thermally unblocked transition inverse to the GT$_+$ resonance. For the  $\overline{\nu}$-absorption, however, the main contribution to the reaction is given by transitions inverse to the GT$_-$ low-energy  strength,
while the "inverse" GT$_-$ resonance is suppressed by the Boltzmann factor in Eq.~\eqref{GT_cr_sect}. Only at $T=1.72$~MeV,  the "inverse" GT$_-$ resonance gives a noticeable contribution to the absorption of low-energy antineutrinos.

As illustrated in the right panels of Fig.~\ref{annb}, the low-energy $^{82}\mathrm{Ge}(\overline{\nu}_e, e^+)$ cross section at finite temperature is also predominantly mediated by GT$_+$ transitions. Although the cross section remains substantially smaller as compared with the other three reactions considered, it demonstrates a strong temperature dependence for antineutrinos with $E_\nu<15$~MeV, i.e. below the ground-state reaction threshold.
Since the downward GT$_+$ transitions are suppressed in  $^{82}\mathrm{Ge}$, the cross section enhancement reflects the thermal unblocking of the upward  GT$_+$ strength. Because of the thermal unblocking, the energy below which the GT$_+$ contribution is more than half of the total cross section shifts to higher values: at $T=0.86$~MeV this energy about 15~MeV, at $T=1.29$~MeV it is about 20~MeV, and at $T=1.72$~MeV it is about 25~MeV. For higher energies forbidden transitions dominate the process like at $T=0$ and the cross-section
depends only weakly on temperature.

\begin{figure}[t!]
 \begin{centering}
\includegraphics[width=\columnwidth, angle=0]{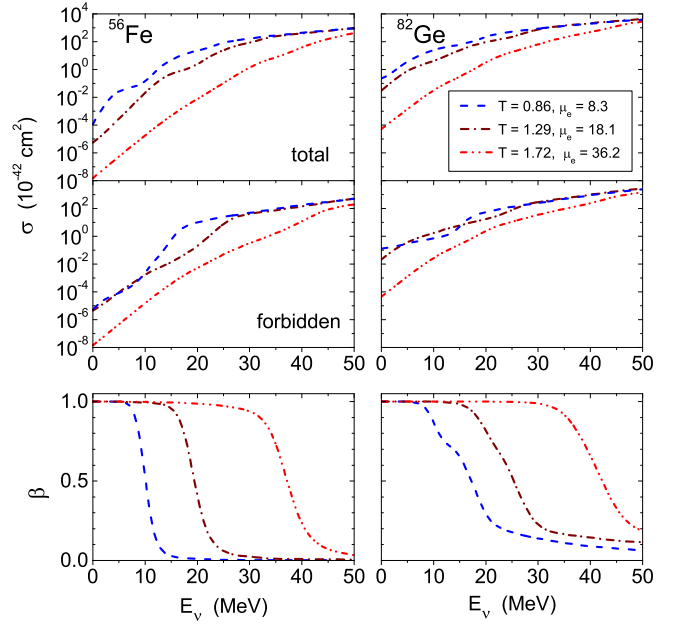}
\caption{(Color online) The same as in Fig.~\ref{nnb}, but now including the outgoing electron blocking} \label{nb}
 \end{centering}
\end{figure}

As was mentioned above, during the core contraction neutrino absorption by nuclei is hindered by Pauli blocking of the phase space for the outgoing electron. To study this effect within the TQRPA, following Ref.~\cite{Sampaio_PLett511},  we have calculated the $\nu$-absorption cross sections by introducing a blocking factor $(1-f(E_e))$ defined at three different sets of temperature  and chemical potential ($\mu_e$ in MeV): $(T,\mu_e)=(0.86, 8.3), (1.29, 18.1),~ \mathrm{and}~ (1.72, 36.2)$.  The results are shown in Fig.~\ref{nb}
for the  $^{56}\mathrm{Fe}({\nu}_e, e^-)$ and  $^{82}\mathrm{Ge}({\nu}_e, e^-)$ reactions. As can be seen, the neutrino absorption cross sections are drastically reduced due to electron blocking in the final state. Moreover, as the chemical potential increases faster than the temperature the cross sections decrease with temperature.
It is obvious that the absorbtion due to de-excitation of thermally excited states is less affected by the Pauli blocking
since the outgoing electron gains energy. This is clearly demonstrated by the plots in the lower panels of Fig.~\ref{nb} where  the relative contribution of negative-energy transitions is shown. We thus conclude that owing to blocking effect negative energy transitions dominate the cross sections up to neutrino energies $E_\nu\gtrsim \mu_e$. Another consequence of the blocking is that downward $1\hbar\omega$ forbidden transitions become important with increasing $\mu_e$. Without Pauli blocking their contribution in negligible due to a small Boltzmann weight. However, as shown in the middle panels of Fig.~\ref{nb}, for  $(T,\mu_e) = (1.29, 18.1)~ \mathrm{and}~ (1.72, 36.2)$ their contribution is comparable or even larger than those of the allowed transitions.  According to our calculations this downward forbidden contribution is mainly due to $0^-,~1^-,2^-$ transitions.

\begin{figure}[t!]
 \begin{centering}
\includegraphics[width=0.9\columnwidth, angle=0]{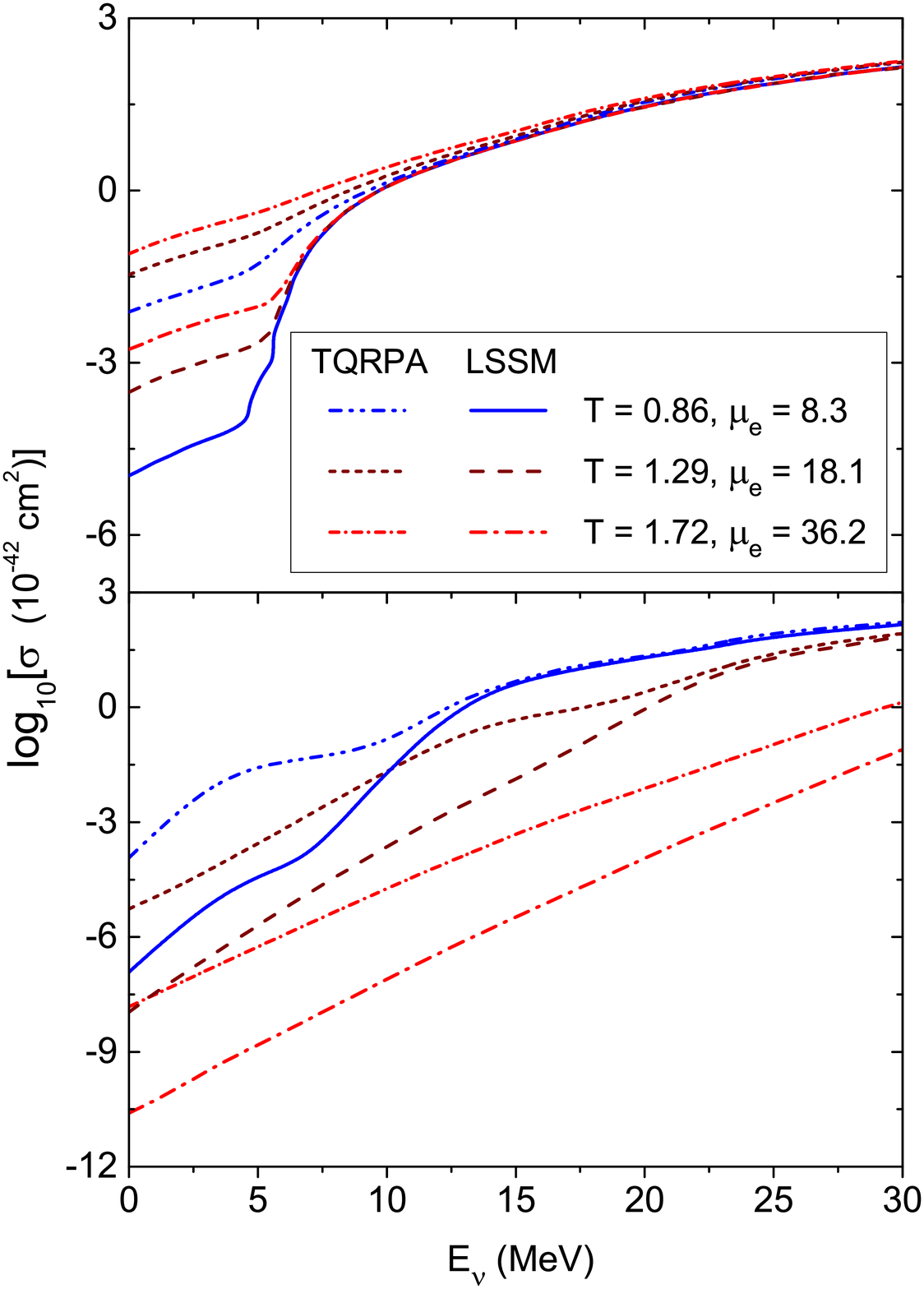}
\caption{(Color online) Comparison of the cross sections of neutrino absorption on the hot nucleus $^{56}$Fe calculated within the TQRPA and the LSSM approach (see Ref.~\cite{Sampaio_PLett511}, Fig.~1).
The cross sections in the upper (lower)  panel are derived without (with) the final-state electron blocking. The respective temperatures and chemical potentials are given in MeV. }
 \label{comparison}
 \end{centering}
\end{figure}

In Fig.~\ref{comparison},  the $^{56}\mathrm{Fe}(\nu_e,e^-)$ cross sections are displayed together with the results of the large-scale shell-model calculations (see Fig.~1 in Ref.~\cite{Sampaio_PLett511}). Even if the general behaviour of the cross sections as functions of the neutrino energy and temperature is in agreement in both approaches, the TQRPA results are noticeable larger than the LSSM ones but the discrepancy reduces with increasing neutrino energies. To understand the cause of the discrepancy, let us consider first the cross sections calculated neglecting the final-state electron blocking. For $E_\nu<5$~MeV the TQRPA cross sections exceed the LSSM values by two to three orders of magnitude. For such low energies, in both approaches,  the neutrino absorption on $^{56}$Fe is dominated by GT$_-$ downward transitions from thermally excited states. The differences in the description of such transitions explain the differences between the TQRPA and shell-model results. As the TQRPA is based on the grand canonical ensemble, the upward and downward GT$_\pm$ strength are connected by the detailed balance.  As shown in Fig.~\ref{partial_T}, the absorption of low-energy neutrinos on $^{56}$Fe is dominated by  the downward GT$_-$ transition inverse to the GT$_+$ resonance. This corresponds to an excitation energy $\omega_i\approx 6$~MeV in the Boltzmann factor of Eq.~\eqref{GT_cr_sect}. Within the shell-model calculations, downward transitions are included by back-resonances, that is by inverting the GT$_+$ strength distribution of the daughter nucleus. In~\cite{Sampaio_PLett511}, the back-resonances are built on the lowest states of $^{56}$Co and the bulk of the back-resonance strength in~$^{56}$Fe is located at an excitation energy of $E_i\approx 7-9$~MeV. This excitation energy is somewhat higher than $\omega_i$. Moreover, within the TQRPA the downward strength concentrates in a single state, while the shell model GT$_+$ strength for   $^{56}$Co is highly fragmented owing to multinucleon correlations~\cite{PLB453}.  It is clear that both these factors suppress the contribution of downward transitions within the LSSM. To see whether the TQRPA reliably predicts the strength of negative-energy transitions, one must go beyond the TQRPA. For a separable residual interaction used here this can be done following the method developed within the QPM, that is by taking phonon coupling into account. On the other hand we should note that due to violation of Brink's hypothesis some back-resonances built on high-lying excited states of  $^{56}$Co may be located at the same energies as those built on the nuclear ground and low-lying states. Due to an increasing density of states, the contribution of such back-resonance states may be substantial and, therefore, their inclusion into the shell-model calculations may  improve the agreement between the TQRPA and shell-model results. For $5<E_\nu<15$~MeV, thermally unblocked low-energy GT$_-$ transitions come into play, and since such transitions do not appear within the shell-model based calculation they also cause the excess of the TQRPA cross sections over the LSSM ones. At $E_\nu>15$~MeV, $\nu$-absorption is dominated by the strong transition involving the GT$_-$ resonance. With increasing $E_\nu$, the cross section becomes insensitive to the energy dependence of the GT distribution and depends only on the total GT$_-$ strength. As a result we observe excellent agreement between the results of both approaches.

Comparing the cross sections calculated with Pauli blocking for the outgoing electron (see the lower panel of Fig.~\ref{comparison}), we note that in~\cite{Sampaio_PLett511}, only allowed GT$_-$ transitions are taken into account when calculating neutrino absorption cross section. As is shown in Fig.~\ref{nb}, if the Pauli blocking  is taken into account, forbidden downward transition become important with increasing temperature and electron chemical potential. It is apparent that such transitions along with the above discussed reasons lead to larger values for the TQRPA cross sections as compared to the shell-model results.

\section{Conclusion}\label{conclusion}

In this work, we have studied thermal effects on the (anti)neutrino absorption for hot nuclei in supernova environments. For this purpose, we have employed the proton-neutron QRPA extended to finite temperatures
within the TFD formalism. As an example, cross sections were calculated for $^{56}$Fe and $^{82}$Ge
in the temperature range from $T=0$ to $1.72$~MeV by taking into account the relevant charge-exchange
transitions $J^\pi=0^\pm,~1^\pm, 2^\pm$, and $3^\pm$.

A detailed analysis of thermal effects was performed for allowed GT transitions which dominate the cross sections
for $E_\nu<30$~MeV neutrinos. Since the TQRPA does not support the Brink hypothesis, new peaks appear in the GT$_\mp$ strength function at finite temperature due to transitions from the excited states. Moreover,  thermal effects shift the GT resonance centroids to lower energies and this effect appears more strongly for the GT$_+$ strength in $^{82}$Ge. The downward transitions from nuclear excited states were included in our calculations through detailed balance. The validity of detailed balance for charge-exchange transitions is a consequence of the grand canonical treatment of hot nuclei.

We have found that thermal effects on the GT strength enhance the absorption cross sections for low-energy (anti)neutrinos by several orders of magnitude. This enhancement is mainly due to increasing contributions of downward transitions from excited states. However, in the supernova environment the electron chemical
potential increases more rapidly than temperature. As a result, if the electron blocking in the final state is taken into account, the neutrino cross sections are drastically reduced.

Although our calculations reveal the same thermal effects as the shell-model calculation,  the calculated low-energy cross sections for $^{56}$Fe exceed the shell-model values by two to three orders of magnitude. One of the possible reason for this discrepancy is that the TQRPA underestimates multinucleon correlations which are
responsible for the GT strength fragmentation. On the other hand, the inclusion of back-resonances built on highly excited daughter states into shell-model calculations may also improve the agreement between the TQRPA and shell-model results.

Since the TQRPA is not restricted by iron-group nuclei, it has some advantages over shell-model calculations.
To enhance its reliability and predictive power several improvements could be made. First of all, to account for multiconfigurational effects, the coupling of thermal charge-exchange phonons with more complex (e.g., two-phonon) configurations should be included into the approach. At zero temperature this problem was considered within the QPM~\cite{Kuzmin_JPG10,Kuzmin_JPG11} by exploiting a separable form of the residual interaction.
The other direction of the improvement is to combine the TQRPA method with self-consistent QRPA calculations based on either the relativistic or Skyrme nuclear energy density functionals. Recently, such calculations were performed at zero temperature~\cite{Lazauskas_NPA792,Paar_PRC84}. With a separable approximation for the Skyrme interaction~\cite{Giai_PRevC57} it will be possible calculate the phonon coupling at $T\ne0$ within a self-consistent theory. This is planned for the future.

\begin{acknowledgments}

We are greatly indebted to Prof. G.~Mart\'{\i}nez-Pinedo for helpful discussions and important comments on this paper. This work was supported by the Heisenberg-Landau Program.

\end{acknowledgments}

\appendix*

\section{}\label{app}

Here we show that the detailed balance condition in the form~\eqref{balance2} can be derived in a model independent way. When considering a grand canonical ensemble of hot nuclei, the probability to find the $i$th excited state of a nucleus with $Z$ protons and $N$ neutrons is given by
\begin{equation}
 P(\varepsilon_i,A^Z_N)=(2J_i+1) \exp\Bigl\{-\frac{\varepsilon_i -\lambda_n N - \lambda_p Z}{T}\Bigr\}\mathcal{Z}^{-1}
\end{equation}
where $\mathcal{Z}$ is the partition function of the grand canonical ensemble and $J_i$ is the angular momentum. Notice that the excitation energies $\varepsilon_i$ are counted from the energy of noninteracting nucleons, i.e., $\varepsilon_0$ is a ground-state binding energy and the chemical potentials $\lambda_{n,p}$ do not include nucleon rest mass. Let us now introduce the  temperature-dependent strength  function for charge-exchange transitions as a thermal average of all transition strengths (probabilities) from states in the parent nucleus to states in the daughter nucleus:
\begin{align}\label{S-}
\Phi^{(\mp)}(E) =& \sum_{Z,N}\sum_{i,f} P(\varepsilon_i,A^Z_N) S^{(\mp)}_{if}\delta(E-Q^{(\mp)}_{if}).
\end{align}
Here $Q^{(\mp)}_{if}=\varepsilon_f-\varepsilon_i\mp\Delta M_{np}$ and
\begin{equation}
 S^{(\mp)}_{if}=\bigl|\langle f,A^{Z\pm1}_{N\mp1}\|\mathcal{T}^{(\mp)}\|i, A^Z_N\rangle\bigr|^2
\end{equation}
is the reduced transition strength between states $i$ and $f$ in the parent and daughter nuclei, respectively. In the above equations the upper sign corresponds to $n\to p$ transitions, while the lower sign refers to $p\to n$ transitions. The transition energy $E$ can be both positive and negative.

For the transition operators $\mathcal{T}^{(-)}$ and $\mathcal{T}^{(+)}$, which differ only by the isospin operator, the respective transition strengths, $S^{(-)}_{if}$ and $S^{(+)}_{fi}$, are connected by detailed balance through
\begin{equation}
  (2J_i+1)S^{(-)}_{if}=(2J_f+1)S^{(+)}_{fi}.
\end{equation}
Combining this result with Eq.~\eqref{S-}, we get the following relationship between the strength functions
for $n\to p$ and $p\to n$ transitions in the thermal grand canonical ensemble
\begin{equation}\label{DB_2}
  \Phi^{(\pm)}(-E)=  \Phi^{(\mp)}(E)\exp\Bigl\{-\frac{E\mp(\Delta\lambda_{np}+\Delta M_{np})}{T}\Bigr\}.
\end{equation}
This relation is exactly the same as derived within the TQRPA approach for charge-changing transitions in hot nuclei.

%\nocite{*}

%\bibliography{dzhioev}

%\end{document}

%

\end{document}